\documentstyle[11pt,epsfig]{article}
\newcommand{\be} {\begin{equation}}
\newcommand{\ee} {\end{equation}}

\begin{document}

\title{\bf A New Look at Low--Temperature Anomalies in Glasses\footnote{To 
appear in: {\em Festk\"orperprobleme/Advances in Solid State 
Physics} {\bf 38} (1998)}}

\author{Reimer K\"uhn and Uta Horstmann\\{}\\
Institut f\"ur Theoretische Physik, Universit\"at Heidelberg\\
Philosophenweg 19, 69120 Heidelberg, Germany}

\date{April 9, 1998}

\maketitle

\begin{abstract}
We review a model--based rather than phenomenological approach to
low--temperature anomalies in glasses. Specifically, we present a
solvable model inspired by spin--glass theory that exhibits both, 
a glassy low--temperature phase, and a collection of double-- and 
single--well configurations in its potential energy landscape. 
The distribution of parameters characterizing the local potential 
energy configurations can be {\em computed\/}, and is found to differ 
from those assumed in the standard tunneling model and its variants. 
Still, low temperature anomalies characteristic of amorphous materials
are reproduced. More importantly perhaps, we obtain a clue to the
universality issue. That is, we are able to distinguish between 
properties which can be expected to be universal and those which 
cannot. Our theory also predicts the existence, under suitable 
circumstances of amorphous phases {\em without\/} low--energy tunneling 
excitations.
\end{abstract}

\section{Introduction}
Ever since the first measurements of Zeller and Pohl \cite{ZePo}
revealed that the specific heat and and the thermal conductivity of glassy 
systems at low temperatures are strikingly different from 
those of crystalline substances, the low temperature physics of glassy 
and amorphous materials has been a subject of intense research efforts,
both experimentally and theoretically.

Specifically, it was found \cite{ZePo} that below approximately 1\,K 
the specific heat of glassy materials scales approximately linearly with 
temperature, $C\sim T$ while the corresponding scaling for the thermal 
conductivity $\kappa$ is approximatly quadratic $\kappa \sim T^2$. Both
findings contrast the $T^3$ behaviour of these quantities in crystals.
Between 1\, K and approximately 20\,K the thermal conductivity exhibits
a plateau and, then continues to rise as the teperature is further 
increased. The specific heat, too, changes its behaviour in the 1--20\,K
regime. It exhibits a bump if displayed in $C/T^3$ plots.

Particularly intriguing is the fact that the anomalous behaviour below
1\,K appears to be universal in the sense that it is shared by virtually
all glassy and amorphous materials, whereas the behaviour between 1 
and 20\,K  seems to display a greater material dependence. 
The universality in the thermal conductivity data actually appears to go 
beyond the level of exponents, in that by a suitable rescaling with a 
constant depending on the Debeye temperature and the the sound velocity, 
thermal conductivity data for various substances can be scaled onto one 
master--curve both below and above, though not on the plateau \cite{FrAn}. 

It should be added that there is also an extensive body of experimental data
characterizing the anomalous response of glassy and amorphous materials to
external probes such as ultrasound or electromagnetic fields. The reader is 
invited to consult the reviews of Hunklinger et al. \cite{Hu} or Phillips
\cite{Ph87} on these matters.

The unusual material properties at low temperatures are usually attributed 
to the existence of a broad range of localized low--energy excitations in 
amorphous systems --- excitations not available in crystalline materials.
At energies below 1\,K, these are believed to be tunneling 
excitations of single particles or (small) groups of particles in double--well
configurations of the potential energy (DWPs). This is the main ingredient of 
the phenomenological so--called standard  tunneling model (STM), independently 
proposed by Phillips \cite{Ph} and by Anderson et al. \cite{An+}. As a second
ingredient of the STM, it is supposed that the local DWPs in amorphous systems
are random, and specific assumptions concerning the distribution of the 
parameters characterising them must be advanced (see below) to describe
the experimental data below 1\,K \cite{Ph,An+}. 

Neither Phillips \cite{Ph} nor Anderson et al. \cite{An+} consider excitations
other than the tunneling excitation in the DWP structures postulated by them
--- their local degrees of freedom are thus true two--level systems (TLSs). 
As a consequence their model is unable to account for the different physics that 
appears in the temperature range between 1 and 20\,K. However, different sets of
low energy excitations in amorphous systems can exist: (i) Higher excitations in 
the abovementioned DWPs --- a possibility that is surely contained 
in the {\em physical\/} picture advanced by Phillips or Anderson et al., but 
was justly considered irrelevant for the very low temperature regime and then 
somehow not reconsidered when problems with the model arose at temperatures
above 1\,K. (ii) Localized vibrations (localized phonons) in soft anharmonic 
single well configurations of the potential energy. The latter have been 
postulated (in addition to tunneling excitations in DWPs) within the, likewise 
phenomenological soft--potential model (SPM) \cite{Kar+,Bu+} to describe the 
physics above 1\,K. Again, local potential energy configurations are supposed 
to be random, and specific assumptions concerning the distributions of the 
parameters characterizing them are required to account for the experimental data
also above 1\,K, such as the crossover to $T^3$--behaviour of the specific heat 
and the plateau in the thermal conductivity \cite{Kar+,Bu+}.

Both, localized soft vibrations \cite{SchoLai} and DWPs \cite{HeuSi} 
responsible for two--level tunneling systems (TLS) have been seen in molecular 
dynamics studies of Lennard Jones glasses. In the case of soft vibrations 
\cite{SchoLai} no attempt, however, was  made to determine the shape of the 
single--well potentials (SWPs) supporting these vibrations, so as to check the 
hypotheses of the SPM. On the other hand, local potential energy configurations 
giving rise to TLSs were analysed within the confines of a generalized SPM, 
\cite{HeuSi} which assumes that locally the potential energy surface (along
some reaction coordinate) can be described by certain fourth order polynomials, 
with coefficients distributed in a specific way. No single--well configurations 
were taken into account, though, to determine the statistics of the coefficients 
\cite{HeuSi}, as in principle they should to make full contact with the 
assumptions of the SPM.

Some time ago, in a lucid critical discussion of the standard theory of TLSs 
in amorphous systems, Yu and Leggett \cite{YuLe} pointed out that there was 
nothing in the STM that could reasonably account for the considerable degree 
of universality observed in amorphous systems. Analogous remarks would mutatis
mutandis apply to the SPM, and it is, of course, mainly related to the 
phenomenological nature of these models. They have to rely on assumptions 
concerning in particular the distributions of parameters characterising local 
potential energy configurations, which --- while plausible in some respects ---
are certainly much less so in others, and are lacking support based on more 
microscopic approaches, such as that of \cite{Gra+} for KBr:KCN mixed crystals.
Neither model, to be sure, accounts for a {\em mechanism\/} that would explain
{\em how\/} the required local potential energy configurations would arise, and
how they would do so with the required statistics.

It is here that our model--based approach to glassy low--temperature
physics \cite{Ku96,KuHo97}, which was started about two years ago, attempts to 
fill a gap.

Our approach takes as its starting point the very observation of 
universality of glassy low--temperature anomalies, which we 
translate into a strategy as follows. Since --- in the light of universality ---
the detailed form of particle--interactions is apparently to a large extent 
immaterial to the phenomena that are to be modeled, we should be justified in 
proposing something like a {\em caricature--glass}. We would like this to be 
understood here in the sense of something as simple as possible as long
as it retains essential properties of glass--forming systems. 
Our success will of course depend on the quality of our understanding as to 
what these essential ingredients  might possibly be. We take them to be (i) 
particles moving in continuous space, subject to (ii) an interaction ({\em any 
interaction!\/}) that gives rise to an amorphous low  temperature phase. Within 
these confines, our choice of the interaction has been mainly guided by the 
demand for simplicity, and analytic tractability. The outcome of these 
deliberations has been a proposal inspired by spin--glass theory 
\cite{Ku96,KuHo97} that describes a glassy material as an anharmonic interacting
particle system with {\em random interactions\/} at the harmonic level, the 
details of which will be specified in Sect. \ref{sec2} below. The interactions 
are chosen in such a way that the model is solvable via mean--field and replica 
techniques well konwn in the theory of spin--glasses \cite{Me+}.

We have organized the remainder of our material as follows. In Sect. \ref{sec2}
we present details of our model, and review main features of its solution within
mean--field theory. We compute its glass transition temperature and exhibit its
phase diagram, featuring ergodic, polarised and glassy phases. Sect. \ref{sec3} 
is devoted to mapping out the potential energy surface of the system in (one
of) its classical glassy ground--states, and to determining the
statistics of its local potential energy configurations. Thermodynamic 
consequences at low temperatures are explored in Sect. \ref{sec4}, where we look
at excitation spectra in local potential energy configurations, at the density
of states, and at the specific heat. Dynamic consequences
 are briefly considered in Sect. \ref{sec5}. We close in 
Sect. \ref{sec6} with a summary and with an outlook on open issues.

\section{Spin--Glass Approach and a Solvable Model}\label{sec2}

We suggest to consider the following Hamiltonian as a candidate for
the description of glassy low--tmeperature properties
\be
{\cal H} = \sum_{i=1}^N \frac{p_i^2}{2 m} + U_{\rm int}(\{v_i\})\ ,
\ee
with an interaction energy given by
\be
U_{\rm int}(\{v_i\}) = -\frac{1}{2} \sum_{i\ne j} J_{ij} v_i v_j + \frac{1}
{\gamma} \sum_i G(v_i)\ ,
\label{uint}
\ee
in which we include an on--site potential of the form
\be
G(v) = \frac{a_2}{2} v^2 + \frac{a_4}{4\,{\rm !}} v^4\ .
\label{gv}
\ee
For the time being we specialize to $a_2=a_4=1$.

The description is in terms of {\em localised\/} degrees of freedom, i.e., the
$v_i$ may be interpreted as deviations of particle positions from a given set 
of reference positions --- in a spirit akin to that used to describe 
dynamical properties of crystalline solids. Thus, we assume the system to be 
already in a solid state, and do not attempt to provide a faithful description 
of the liquid phase.\footnote{Note that we are writing down what appears to be
a scalar model here. The 3-$D$ nature of particle coordinates is, however, 
included in our description, if we choose to consecutively enumerate the 
cartesian components of the particles, in which case $N$ has to be read as 
three times the nuber of particles of the system.}

We propose to model the glassy aspects by taking the interaction at the harmonic
level, that is,  the $J_{ij}$, to be random, so that the reference positions
will generally be unstable at the harmonic level. This is why we have added 
the on-site potentials $G(v)$, namely to stabilize the system as a whole. 
By including in $G$ a harmonic term, we can use the parameter $\gamma$ to 
tune the number of modes that actually {\em are\/} unstable at the harmonic 
level of description. Apart from its role in this `tuning' aspect, we have 
opted for the simplest choice that achieves the required stabilization.

We believe our opting for random interactions to be justified by appeal to 
universality (the detailed form of the interactions cannot be crucial for the 
emergence of glassy low--temperature anomalies), and by the observation made 
in recent years of the fundmental similarity between truly quenched disorder 
and so--called self--induced quenched disorder \cite{Mez++}.

We choose the random interactions in such a way that the system is amenable to 
analysis within replica mean--field theory, well known from the theory
of spin--glasses\cite{Me+}. Specifically, we choose them to be Gaussian 
random variables with $\overline{J_{ij}} = J_0/N$ and $\overline{J_{ij}^2}
-\overline{J_{ij}}^2 =1/N$.
The harmonic part of $U_{\rm int}(\{v_i\})$ is thus reminiscent of the SK
spin--glass \cite{SK}, except for the fact that we are dealing here with 
continouus degrees of freedoms rather than with classical Ising spins. 
Once more, by appeal to universality, and by noting that we do not intend 
to describe correlations at a critical point, it is not unreasonable to argue
that a mean--field description should be sufficient to reveal the essential 
physics we are after.

Let us mention for completeness that models with an interaction energy of the 
form (\ref{uint}) were considered before in an entirely different field, 
namely in the context of analog--neuron systems and networks of operational
amplifiers \cite{Ho,KuBoe}. Incidentally, this connection opens up an 
unexpected but rather fascinating perspective of studying glassy dynamics
by emulating it directly in highly integrated circuitry, albeit perhaps not 
in the regime of low temperatures, where quantum effects are important.

At the classical level of description, all interesting information about
the thermodynamics of the system is obtained from the configurational free
energy
\be
f_N(\beta) = -(\beta N)^{-1} \ln \int \prod_i d v_i \exp[-\beta U_{\rm 
int}(\{v_i\})]\ .
\label{fnbeta}
\ee
The free energy has to be averaged over the ensemble of possible relisations
of the $J_{ij}$ configurations so as to get {\em typical\/} results. This 
is accomplished by means of the so--called replica method, well
known in the theory of disorderd systems (see, e.g. \cite{Me+}). We will
not here reproduce the calculations, but merely quote the result. The quenched 
free energy  --- that is, the disorder--average of (\ref{fnbeta}) --- is 
represented as the limit $f(\beta) = \lim_{n\to 0} f_n(\beta)$, with
\begin{eqnarray}
n f_n(\beta) & = &\frac{1}{2} J_0 \sum_a p_a^2 + \frac{1}{4}\beta \sum_{a,b} 
q_{ab}^2 \nonumber \\
& - & \beta^{-1} \ln \int \prod_a d v^a \exp\big[-\beta U_{\rm eff} (\{v^a\})
\big]\ .
\label{fn}
\end{eqnarray}
Here
\be
U_{\rm eff}  = - J_0 \sum_a p_a v^a - \frac{\beta}{2} \sum_{a,b} 
q_{ab}v^a v^b  + \frac {1}{\gamma}\sum_a G(v^a)
\label{ueff}
\ee
is a replicated single--site potential and the order parameters $p_a =N^{-1} 
\sum_i {\langle v_i^a\rangle}$ and $q_{ab} =N^{-1} \sum_i {\langle v_i^a v_i^b\rangle}$ are determined from the fixed point equations
\begin{eqnarray}
p_a & = &\langle v^a \rangle \quad ,\ a=1,\dots ,n \\
q_{ab} & = &\langle v^a v^b \rangle \quad ,\ a,b = 1,\dots ,n\ ,
\label{fpe}
\end{eqnarray}
where $\langle\dots\rangle$ denotes a Gibbs average corresponding to the 
replica potential (\ref{ueff}). The limit $n\to 0$ is eventually to be 
taken in these equations.

\begin{figure}[t]
{\centering 
\epsfig{file=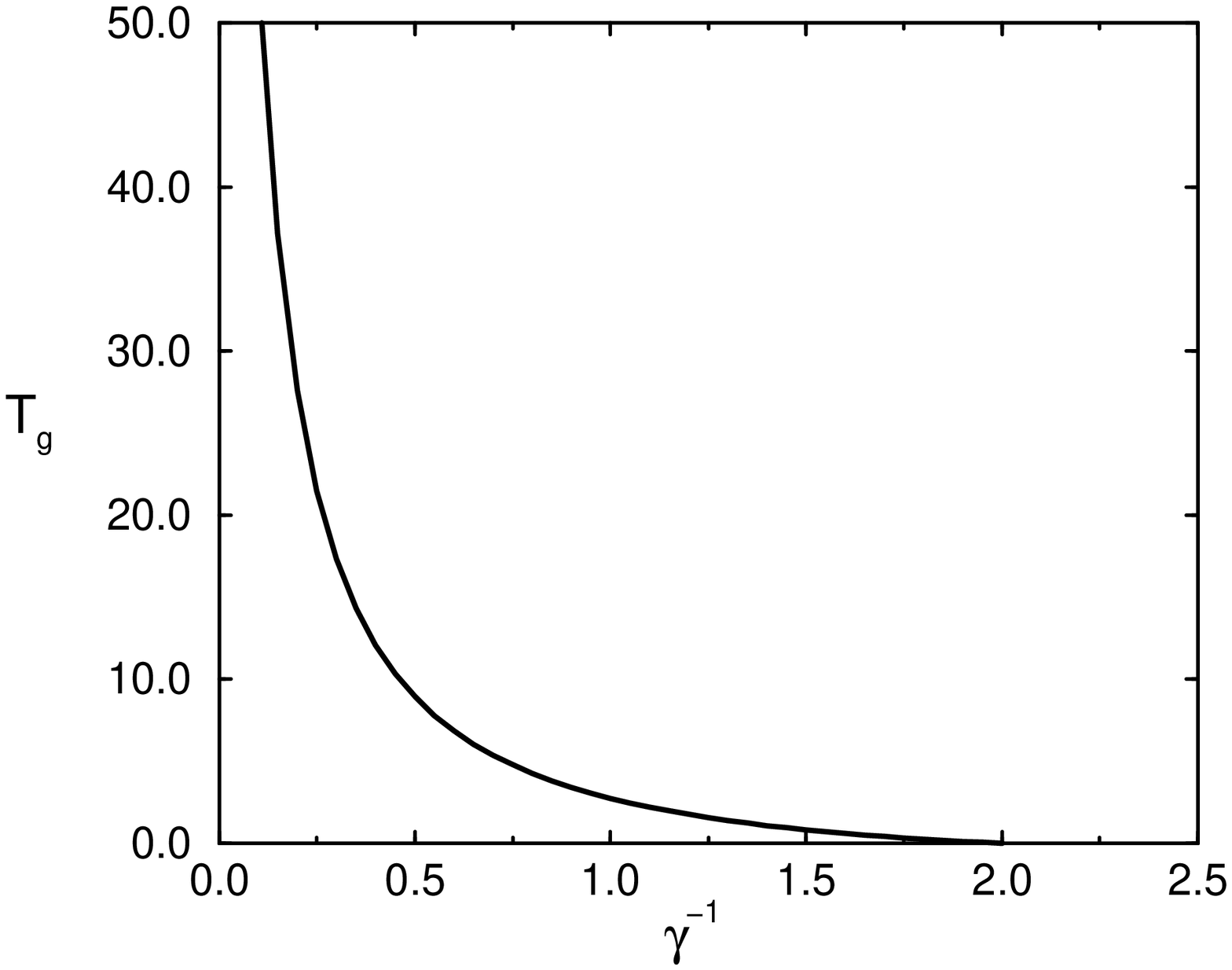, width=7cm, height=6cm}  
\hfill{}
\epsfig{file=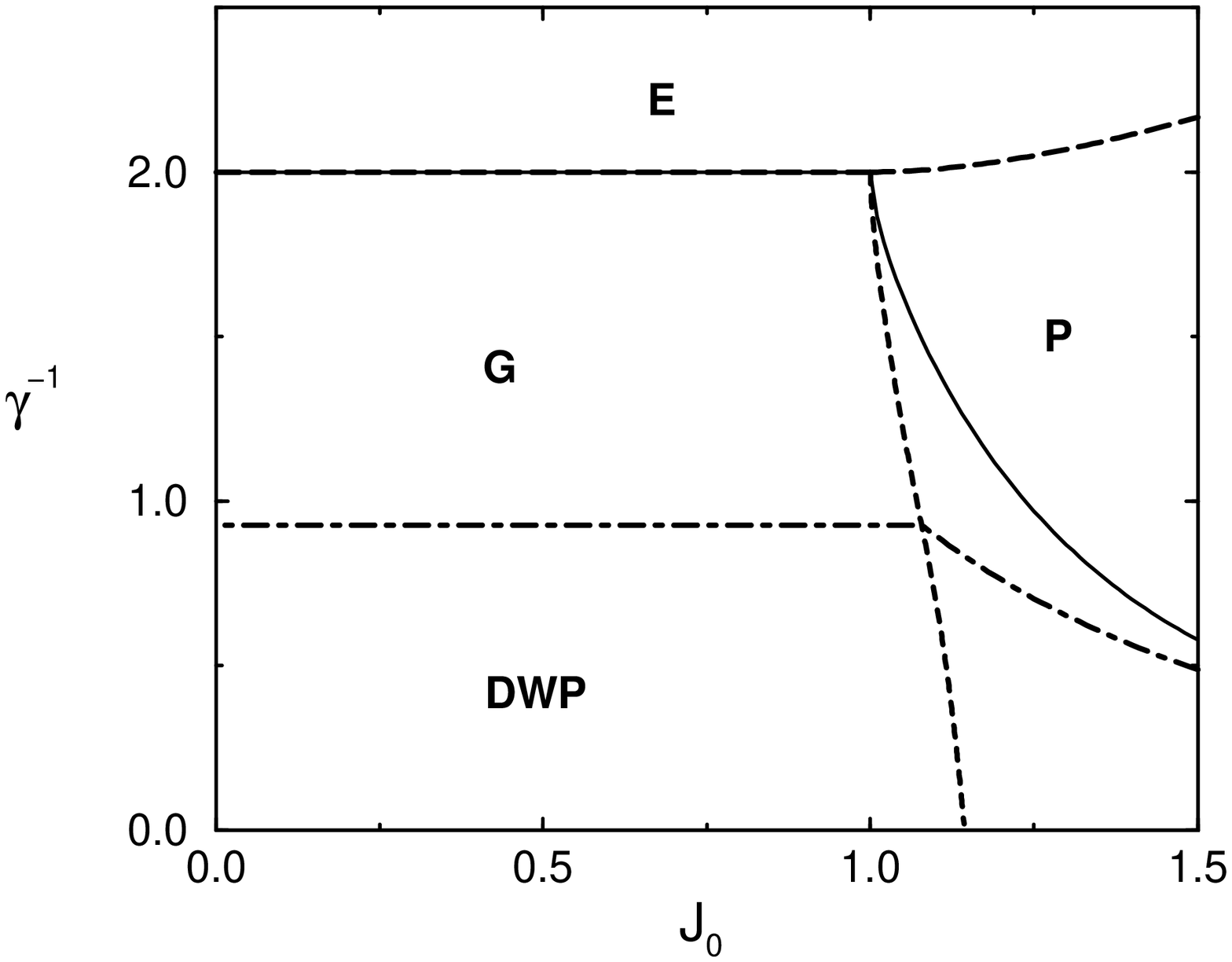, width=7cm, height=6cm} 
\par}
\caption[fig1]{{\bf(a):} Glass temperature $T_g$ as a function of $\gamma$ for 
$J_0=0$. {\bf(b):} $T=0$ phase diagram in RS. $E$ denotes the $T=0$--limit of 
the ergodic phase, $G$ is the glassy phase, and $P$ is a phase with macroscopic 
polarization. The full line is the AT line (see text). Below the dot--dashed 
line is the region with DWPs, as described in detail in Sect. \ref{sec3}. 
The short--dashed line separates the glassy phase $G$ from the phase $P$ with 
macroscopic polarization. From \cite{KuHo97}}
\label{fig1}
\end{figure}

Eqs. (\ref{fn})--(\ref{fpe}) were analysed in the replica symmetric (RS) 
\cite{Ku96} and the 1st step replica--symmetry breaking (1RSB) approximations
\cite{KuHo97}. In RS one assumes $p_a = p$ for the `polarization', 
and $q_{aa}=\hat q$ and  $q_{ab}=q$ for $a\ne b$ for the entries of the 
Edwards-Anderson matrix. These are given as solutions of
\begin{eqnarray}
p  = \langle\, \langle v\rangle\, \rangle_z\ , \  \
C  = \Big\langle\, \frac{1}{\sqrt q}\frac{d\,\langle v\rangle}{d z}\, 
\Big\rangle_z\ , \ \
q   =   \langle\, \langle v\rangle^2\, \rangle_z\ .
\label{fpers}
\end{eqnarray}
Here  $C=\beta (\hat q - q)$, and $\langle\dots\rangle_z$ denotes an average 
over a standard Gaussian $z$ while $\langle\dots\rangle$ is now a thermal 
average corresponding to the RS single--site potential
\be
U_{\rm eff}(v)  = - h_{\rm eff}\, v -\frac{1}{2} C v^2 + \frac{1}{\gamma} G(v)\ ,
\label{urs}
\ee
with 
\be
h_{\rm eff} = h_{\rm RS} = J_0 p + \sqrt{q}\, z
\label{hrs}
\ee 
It turns out that our model exhibits a transition from an ergodic phase
with $p=q=0$ to a glassy phase with $q\ne 0$ at some temperature $T_g$ 
which depends on $J_0$ and $\gamma$. For sufficiently large $J_0$, the 
transition is into a polarized phase with $p\ne 0$. The glass--transition
temperature as a function of $\gamma$ is shown in Fig. \ref{fig1}a. In
Fig. \ref{fig1}b, we exhibit also the $T=0$ limit of the phase diagram, which
will become of importance for the discussion of of the potential energy
surface of the model in Sect. \ref{sec3} below.

The assumption of RS is not justified at low temperatures and large $\gamma$,
where replica symmetry breaking (RSB) --- an instability that breaks the 
permutation invariance among the $n$ replica --- is expected to occur. The 
location of the instability against RSB is given by the de Almeida--Thouless 
(AT) criterion \cite{AT} $1=\beta^2 \, 
\left \langle\,(\,\langle v^2\rangle - \langle v\rangle^2\,)^2\, \right 
\rangle_z$. In a one-step replica-symmetry breaking (1RSB) approximation, which 
is believed to constitute a major step towards the full solution, 
$U_{\rm eff}(v)$ is of the same form (\ref{urs}), however with $h_{\rm RS}$ 
replaced by $h_{\rm 1RSB} =  J_0 p + \sqrt{q_0}\, z_0 + \sqrt{q_1-q_0}\, z_1$, 
and $C$ by $C=\beta (\hat q - q_1)$, where we use standard notation 
\cite{Me+,SK} for the entries of the $q$--matrix. Along with 
a so--called partitioning parameter $m$, they are determined from a more 
complicated set of fixed point equations \cite{Me+,HoKu}. In 1RSB, $z_0$ is a 
Gaussian, whereas the distribution of $z_1$ is more complicated.

In the following section we will set out to demonstrate that the mean--field 
solution of the model we have obtained here does in fact also contain information
about the potential energy surface of the model, the statistics of which is
generally believed to be crucial for the emergence of glassy low--temperature
anomalies.

\section{Mapping Out the Potential Energy Surface}\label{sec3}

To map out the potential energy surface of the system, we observe that all 
that mean field theory is about, is to represent the interaction energy of
the system as a sum of effictive independent single site potential energies
\be
U_{\rm int}(\{v_i\}) \longrightarrow \sum_i U_{\rm eff}(v_i)\ ,
\label{sumui}
\ee
self--consistently to be determined in such a way as to get thermal averages
correct. In the context of non-random systems, the most famous example is
the Curie-Weiss model of ferromagnetism, where the effective single site
potential is the Zeemann energy of a spin in an effective (non-fluctuating)
local- or mean--field determined by its neighbours.

In our case, we obtain the sum (\ref{sumui}) of independent single site 
potentials $U_{\rm eff}(v_i)$, which contain random parameters. Replica 
theory achieves nothing but a selfconsistent determination of the 
distribution of these random parameters. Within the RS approximation these 
effective local single site potentials are given by (\ref{urs}),(\ref{hrs}), 
containing a single random parameter, viz the Gaussian distributed effective 
fields $h_{\rm RS}$, having mean $J_0 p$ and variance $q$. The parameters
of $p$, $q$ and $C$  characterizing the Gaussian ensemble of single-site 
potentials $U_{\rm eff}(v)$ are determined self--consistently through 
(\ref{fpers})--(\ref{hrs}). In 1RSB, the ensemble of single site potentials
is again of the form (\ref{urs}), but now with $h_{\rm RS}$ replaced by the
more complicated $h_{\rm 1RSB}$ and $C$ by $C=\beta (\hat q - q_1)$ as noted
above. Again there is only a single random parameter, a locally varying effective
random field. This feature is easily seen to persist at all levels of replica
symmetry breaking.

There is one additional ingredient, namely we take the the $T=0$ limit of the
theory to select one of the (possibly many) collective classical glassy 
groundstate configurations.

In terms of the mean field solution, we thus have a representation of the
glassy potential energy landscape as an ensemble of randomly varying effective
single--site potentials. Their statistics is determined by the statistics
of the effective fields $h_{\rm RS}$ and $h_{\rm 1RSB}$ in the RS and the 
1RSB approximation, respectively.

Now note in particular, that the initial stabilising concave upward single
site potential $G(v)$ also appears in $U_{\rm eff}(v)$, but there is now
an additional harmonic term, viz. $-\frac{1}{2} C v^2$ --- entirely of 
collective origin --- that renormalises the effective total harmonic restoring 
force. For
\be
C - \frac{1}{\gamma} > 0
\ee
the total harmonic contribution to $U_{\rm eff}(v)$ becomes convex downward
near the origin $v=0$, so that for sufficiently small $h_{\rm RS/1RSB}$
the effective single site potential $U_{\rm eff}(v)$ attains a DWP--form,
which is of collective origin! The region in parameter space where DWPs exist
and its boundary are indicated in the $T=0$ phase diagram in Fig. \ref{fig1}b.

We are now able to make contact with the phenomenology of glassy 
low--temperature anomalies, and to compare with the ideas underlying their 
explanation within the STM \cite{An+,Ph} or the SPM \cite{Kar+}. The main 
difference and, we believe, important new perspective added by our approach
is that the appearance of DWPs and the statistics of local potential energy
configuraqtions need not be hypothesised from the outset, but rather that it 
emerges as a collective effect, originating in the frustrated interactions 
$J_{ij}$, if the temperature is sufficiently low and the parameter $\gamma$ 
sufficiently large. Moreover, the statistics of the local potential energy 
configurations has been rendered part of the world of the {\em computable}. 
It follows directly from the statistics of the  $h_{\rm RS/1RSB}$.

To explore the consequences for the low temperature behaviour of glasses 
quantitatively, one solves the quantum mechanical problem for $U_{\rm eff}(v_i)$
described be the Hamiltonian
\be
H_{\rm eff} = \frac{p^2}{2 m} + U_{\rm eff}(v)
\label{Heff}
\ee
with $U_{\rm eff}(v)$ given by (\ref{urs}),(\ref{hrs}) in the RS approximation,
or by the corresponding expression valid for the 1RSB (or higher order
RSB) ans\"atze.

In the STM one approximates the tunnel-splitting $\varepsilon$ of the 
ground-state in the DWP case to be given in terms of the asymmetry $\Delta$ of
the DWP and the tunneling--amplitude $\Delta_0$ by the TLS-result
\be
\varepsilon = \sqrt{\Delta^2 + \Delta_0^2}\ ,
\ee
the value of $\Delta_0$ being related with the barrier--height $V$ between 
adjacent minima and the distance $d$ between them. In WKB--approximation one has
$\Delta_0 \simeq \hbar\omega_0 \exp(-\lambda)$, with $\lambda=\frac{d}{2}
(2 m V /\hbar^2)^{1/2}$, where $\omega_0$ is a characteristic frequency (of the 
order of the frequency of harmonic oscillations in the two wells forming the 
double well structure), and $m$ the effective mass of the tunneling particle.
Moreover, a specific assumption is advanced concerning the distribution 
of asymmetries $\Delta$, and tunneling--matrix elements $\Delta_0$, namely, in 
terms of $\Delta$ and $\lambda$ one assumes $P(\Delta,\lambda) \simeq 
{\rm const.}$ Note that this assumption implies in particular also that these 
quantities are {\em uncorrelated}. 

\begin{figure}
{\centering 
\epsfig{file=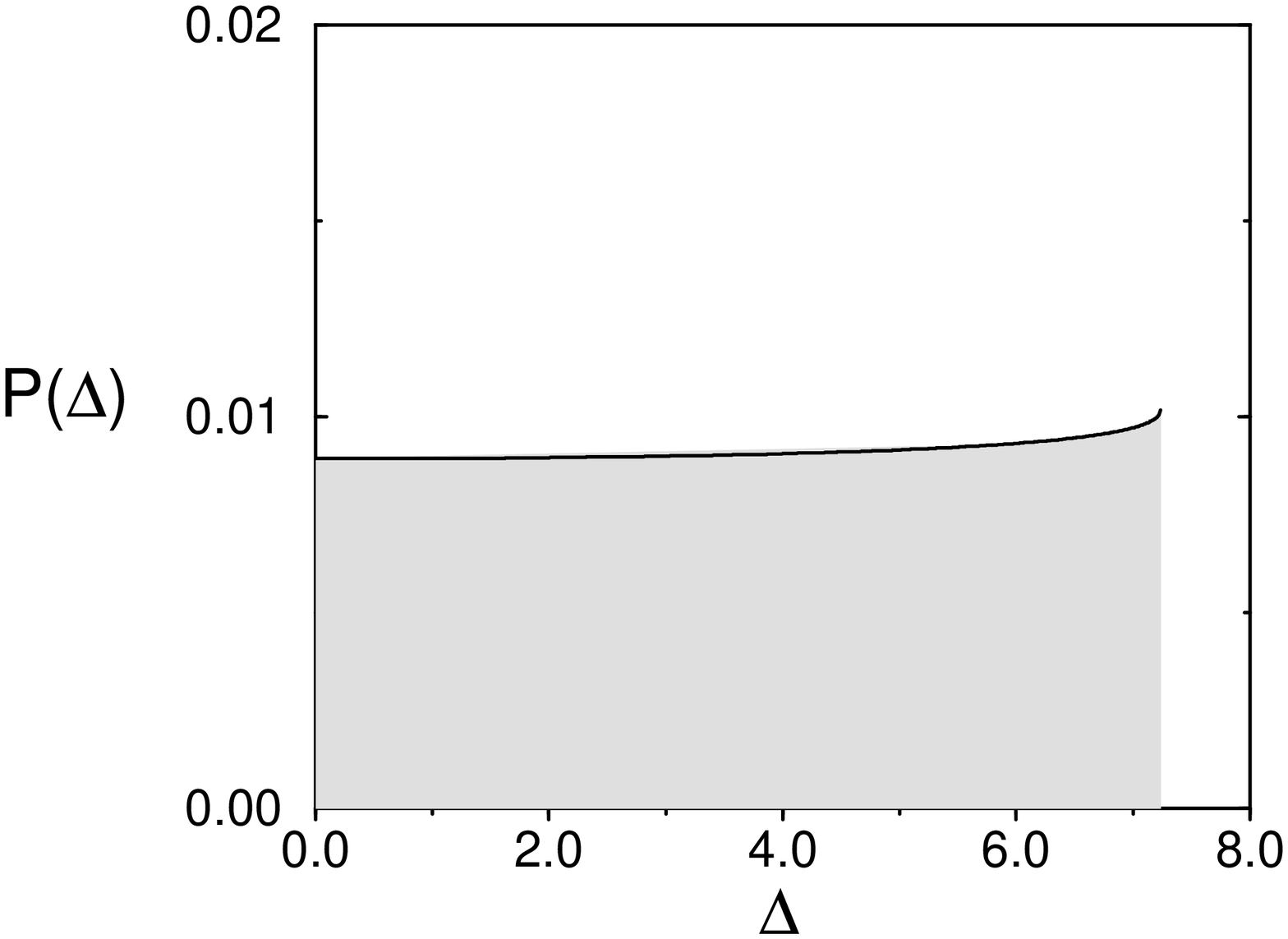, width=7cm, height=6cm}  
\hfill{}
\epsfig{file=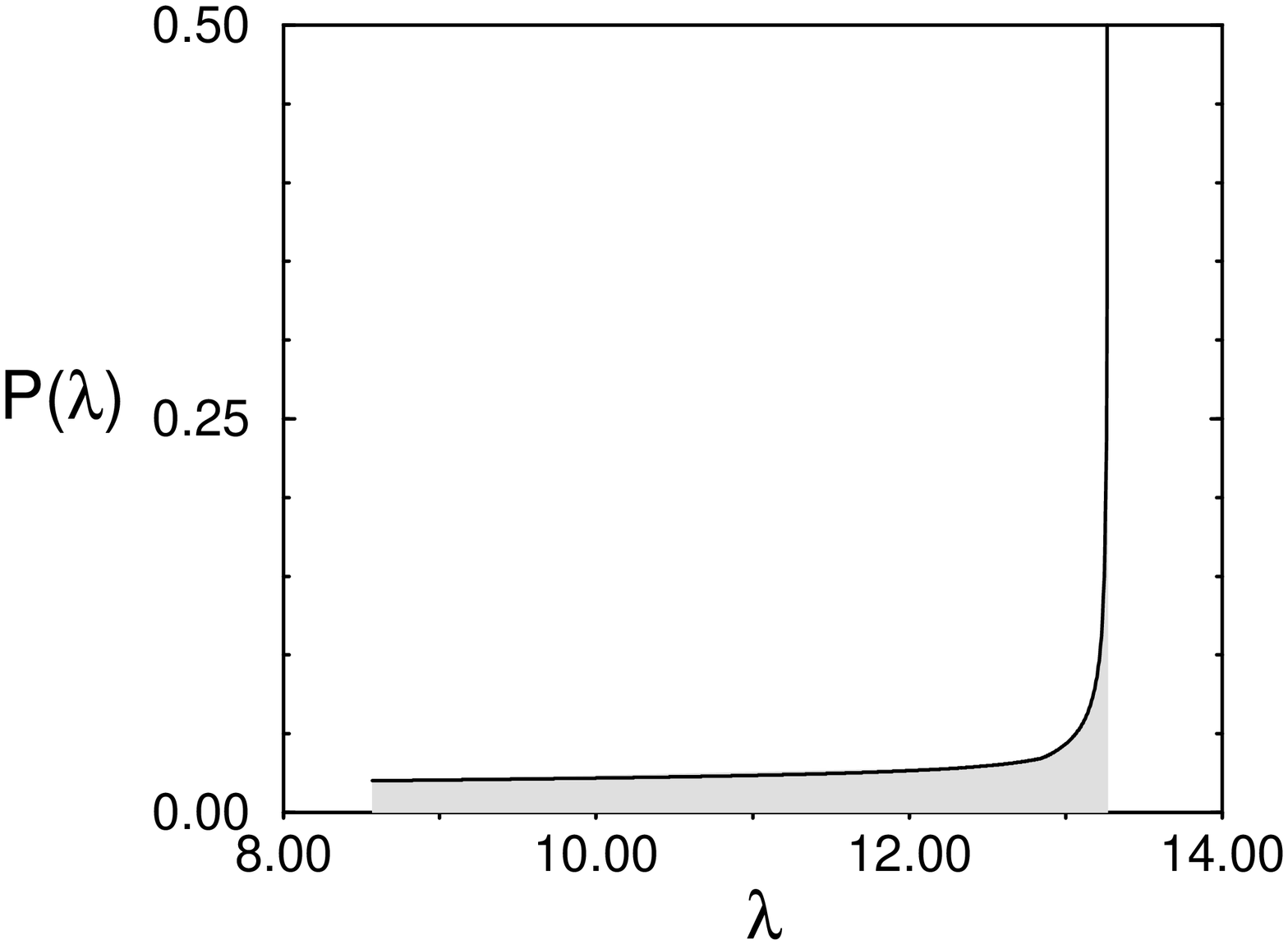, width=7cm, height=6cm} 
\par}
\caption[fig2]{Distributions ${\cal P}(\Delta)$ (for positive $\Delta$), and 
${\cal P}(\lambda)$ in RS. Here $\gamma = 4$, while $J_0=0$. After \cite{Ku96}}
\label{fig2}
\end{figure}

We can check these assumtions against the results from our model \cite{Ku96}. 
Obviously, since the values of both $\Delta$ and $\lambda$ are determined 
by a single random variable, viz. $h_{\rm RS/1RSB}$, these quantities must
be {\em perfectly correlated\/} in our model, in contrast to the assumptions of 
the STM, although their distributions separately are indeed rather flat, except 
for anomalies at their upper end (see Fig. \ref{fig2}) where due to large 
$h_{\rm RS/1RSB}$ DWPs cease to exist.

The perfect correlation can be weakended, but {\em not\/} eliminated, by 
allowing {\em local randomness}, that is, by making the
coefficients $a_2$ and $a_4$ in (\ref{gv}) vary randomly from site to
site. A randomly varying $a_2$ is, in fact, a natural choice in our 
approach, since the $a_2$ contribution to $U_{\rm int}$ should 
be considered part of the harmonic level, which we had assumed to be random 
to begin with. Indeed, by allowing this kind of randomness, our results can
be brought into better agreement with the experimental situation (see below).
As to $a_4$, we have observed that taking it to be random (within limits
of course) has no dramatic effect, so we adhere to it remaining non--random, 
serving solely, as it should, its stabilising purpose. We should add that our 
model remains solvable with these modifications. Outer averages in the fixed 
point equations simply have to be read as implying an additional average over 
the local randomness.

As a final unexpected result, note that our theory predicts the existence under
suitable circumstances of an amorphous phase without DWPs, and hence without
low--energy tunneling excitations (see Fig. \ref{fig1}b). Until very recently, 
this would perhaps have universally been considered rather a surprise. However,
recent internal friction measurements in specially prepared amorphous Si:H 
films by Liu et al. \cite{Liu+97} indicate that such a possibility has to be 
seriously taken into account.

\section{Thermodynamics}\label{sec4}

Information about the thermodynamics of the system at low temperatures, where
it is dominated by quantum effects, is obtained by studying the excitation 
spectra of the local Hamiltonians (\ref{Heff}), and by using this information
--- avaraging it over the randomness characterizing the ensemble of these 
local Hamiltonians, i.e. over the $h_{\rm RS/1RSB}$--distributions (and
possibly the $a_2$--distribution) --- to compute densities of state and 
thermodynamic functions the usual way.

\begin{figure}
{\centering 
\epsfig{file=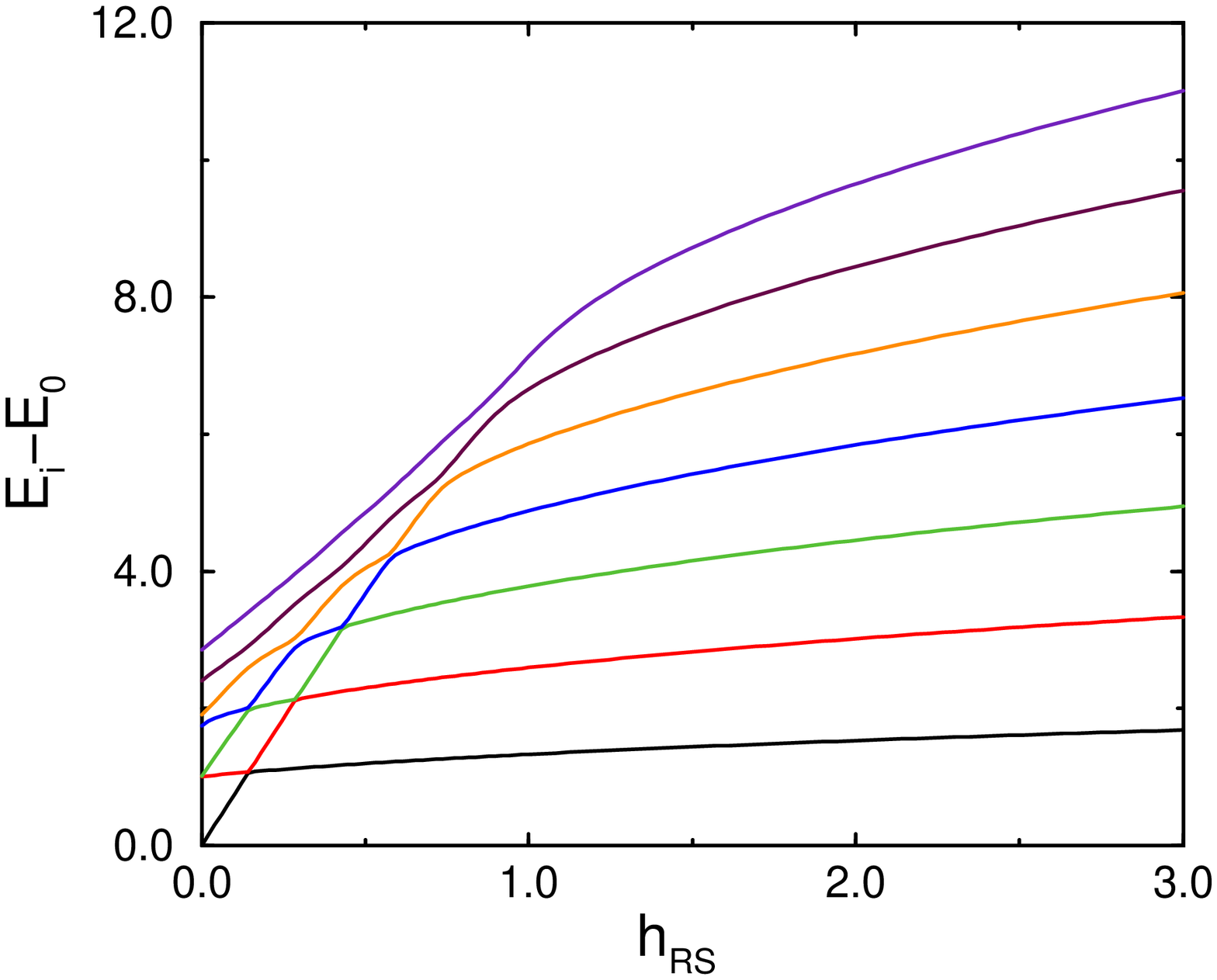, width=7cm, height=6cm}  
\hfill{}
\epsfig{file=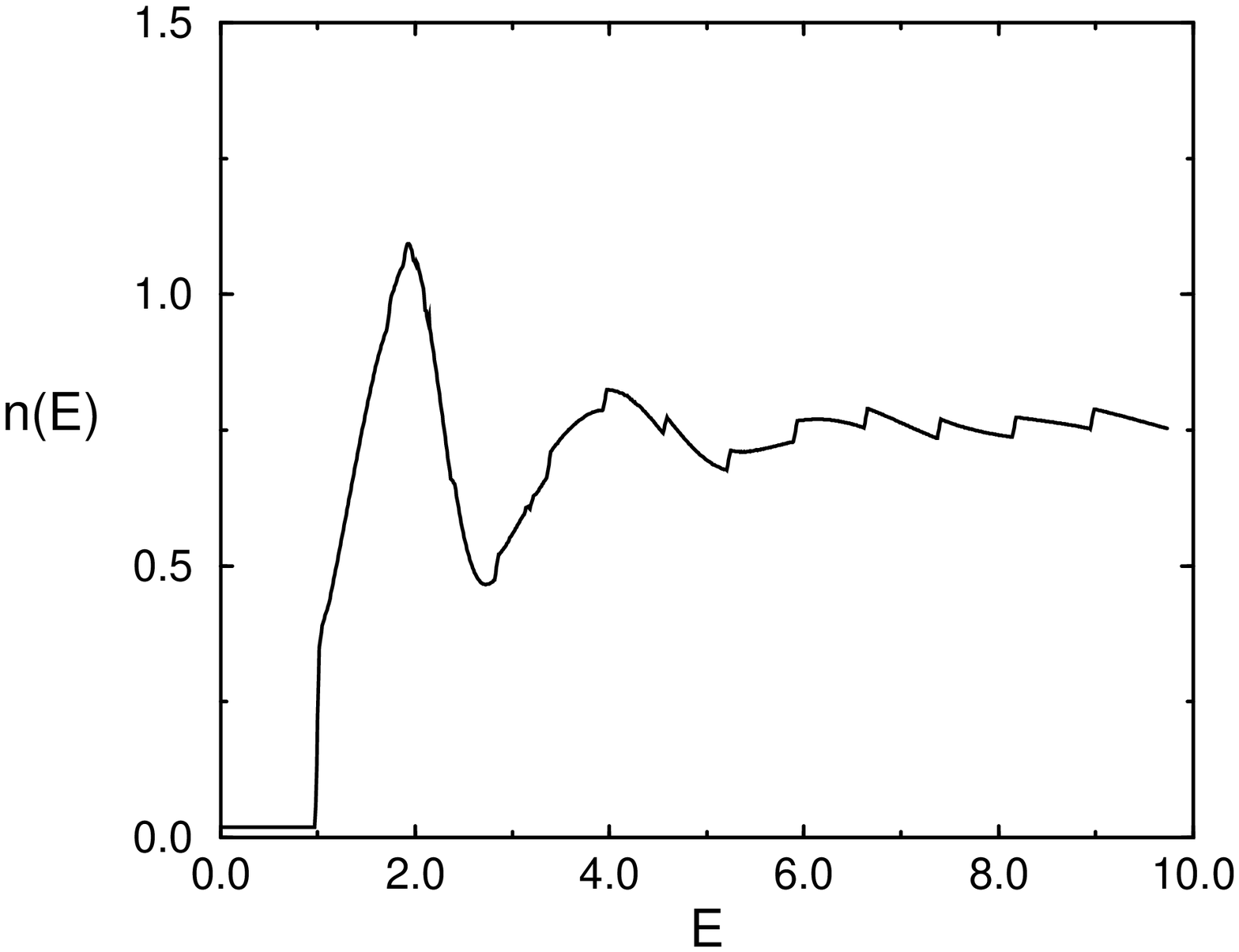, width=7cm, height=6cm} 
\par}
\caption[fig3]{{\bf a):} Excitation spectrum as a function of $h_{\rm RS}$. 
$\gamma = 4$ and $J_0=0$. are as in Fig. \ref{fig2}. {\bf b):} Density of
states in RS, for parameters as in {\bf (a)}.}
\label{fig3}
\end{figure}

In Fig. \ref{fig3}a we exhibit a typical excitation spectrum, as it varies
with the value of $h_{\rm eff}$. The salient features are two kinds of branches,
steeper and less steep ones. The former are related to tunneling (inter--well) 
excitations whereas the the latter are of intra--well type. That is, the latter
would also exist if the wells were not communicating via a barrier of finite
height and width.  Note that these families of levels do not cross as might 
appear at first sight. Rather, there are avoided `would-be' level--crossings 
precisely {\em due} to the tunneling mechanism. At at energies higher than the
barrier or at large $h_{\rm RS}$ the avoided crossing--pattern disappears, 
because there is no longer a barrier to be tunneled through. 

The corresponding density of states (DOS) (averaged over the ensemble) is 
displayed in Fig. \ref{fig3}b. Note its constant value at energies lower than 
that of the first intra--well excitation. It arises because, (i) apart from the 
immediate vicinity of its lower cutoff $\Delta_0$, the level--splitting
of the ground--state scales linearly with the asymmetry $\Delta$, which in turn 
is proportional to $h_{\rm eff}$, and (ii) because the distribution of 
$h_{\rm eff}$ is virtually flat in this regime.

\begin{figure}[t]
{\centering 
\epsfig{file=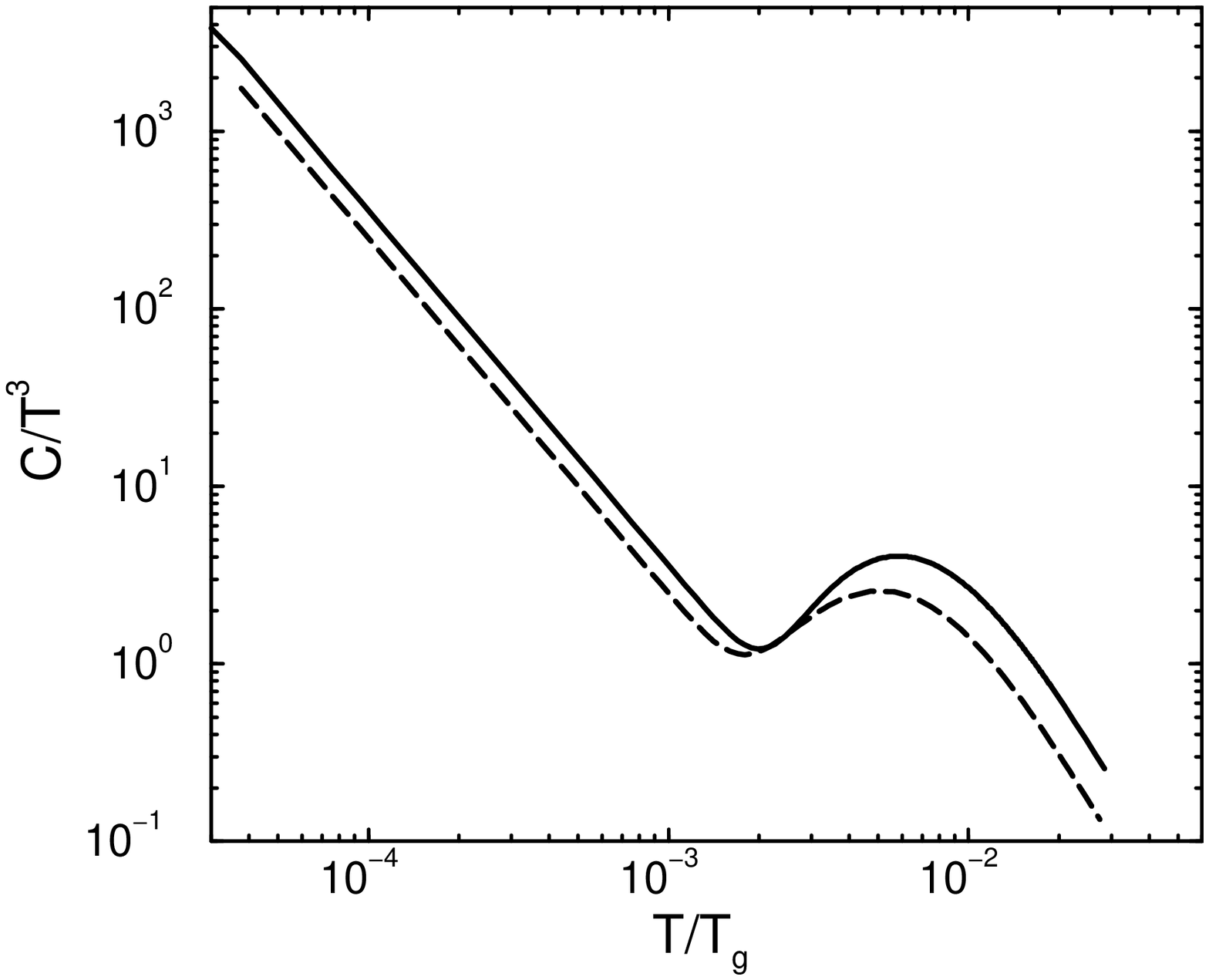, width=7cm, height=6cm}  
\hfill{}
\epsfig{file=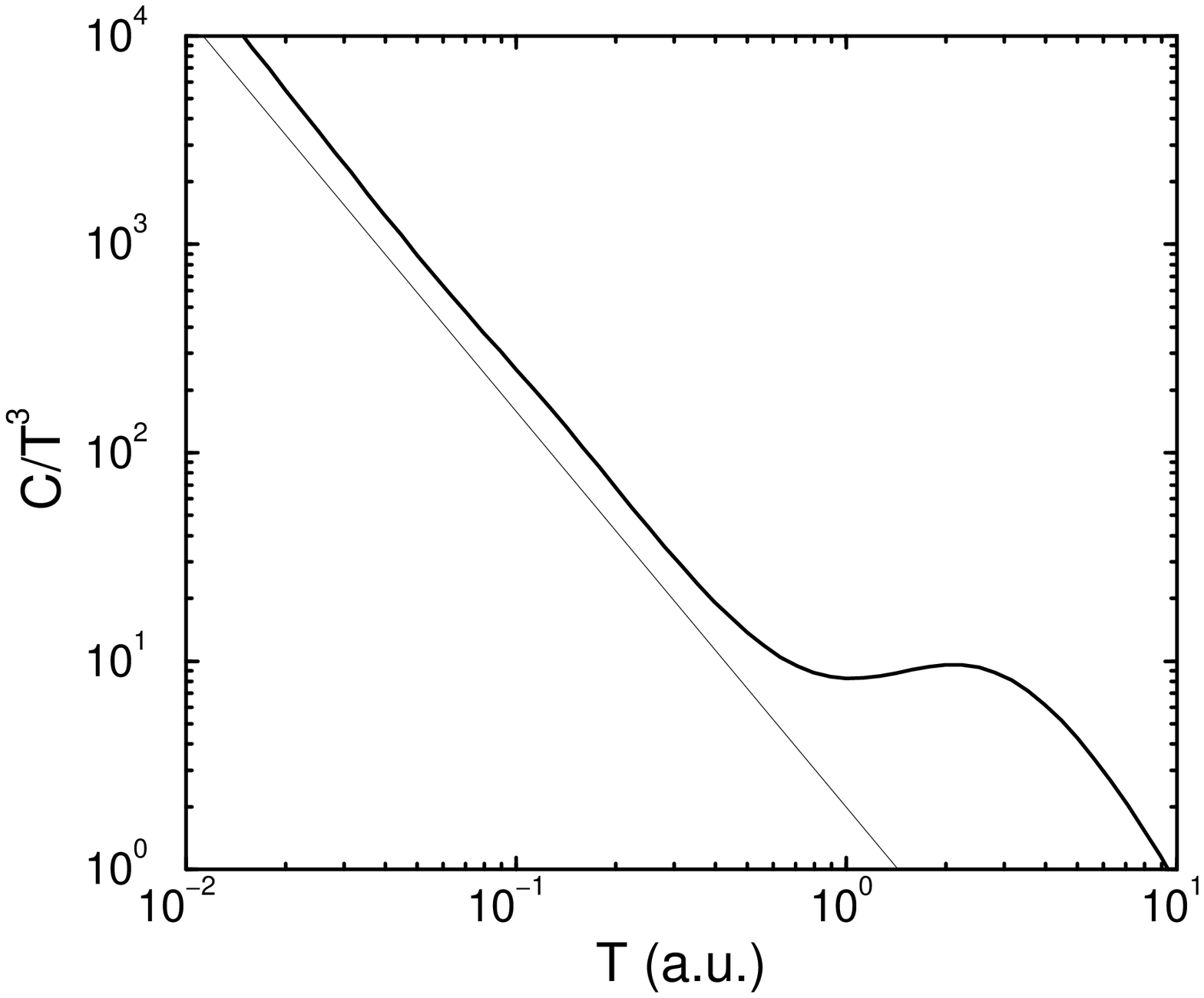, width=7cm, height=6cm} 
\par}
\caption[fig4]{Specific heat at low temperatures {\bf a):} without local 
disorder (full line RS, dashed line 1RSB) , {\bf b):} with $\gamma=1$, and
$a_2$ a Gaussian of mean 1 and variance $\sigma=5$ in RS. Other parameters 
are as before. The thin line corresponds to $C\sim T^{1.1}$.}
\label{fig4}
\end{figure}

This energy range is to be associated with the {\em universal\/} glassy 
low--temperature anomalies below 1\,K! Its properties are mainly driven by 
the distribution of effective fields $h_{\rm eff}$ and are thus entirely of 
collective origin. In particular it is related to the temperature range in 
which the specific heat scales linearly with temperature (see Fig. \ref{fig4}).

By contrast, the energies of the intra--well excitations are 
much less sensitive to the value of the effective field, hence less steep in 
the excitation spectrum if plotted vs. $h_{\rm eff}$. These give rise to 
the strong increase of the DOS at the value of the first intra--well excitation 
(the contribution to the DOS being proportional to the {\em inverse} slope of 
the level as a function of $h_{\rm eff}$). The intra--well excitations 
depend more strongly on the shape of the {\em local\/} stabilizing potentials 
$G(v)$, hence they cannot be of collective origin --- basically they turn out 
where we have chosen them to be, with or without local randomness in the
$G(v)$. So the physics that depends on these features, namely the bump in
$C(T)/T^3$ and the plateau in the thermal conductivity, cannot be expected 
to be universal. 

This separation within our approach of aspects that can be expected to be 
universal from those which cannot is in perfect qualitative agreement with 
the experimental situation. It is not unreasonable to suppose that such a
distinction between local and global contributions to $U_{\rm int}$ can be
made for real materials as well, the contribution of the long--range global
one being for instance of elastic nature (and well known to give rise to 
frustration).

Note that our representation of local potential energy configurations has
finally turned out to be similar to that proposed within the SPM \cite{Kar+,Bu+}
with some marked differences in the details: (i) Within our approach the very
occurrence of DWPs is of identifiable collective origin, and so is the 
distribution of asymmetries. (ii) {\em No\/} analogous mechanism is available
for determining (distribution of) the parameters entering $G(v)$, which are
mainly responsible for the distribution of barrier heights and the nature of
(single- and) intra-well excitations which determine the material--specific
properties. From our vantage point, therefore, one should not even attempt
at proposing for these something universally valid for all glassy materials.

There is some influence of local disorder on the behaviour of 
the low $T$ specific heat also in the `universal' regime changing it to 
slightly superlinear, with exponents depending on the nature of the
distribution. This feature, too, agrees nicely with experimental findings,
and it might be used to give some handle at fixing the distribution of
local parameters for specific materials.

Concerning universality, Yu and Leggett supposed that it could result only due
to a sufficiently long range interaction between TLSs, an idea which has since
the been pursued in a series of papers by Burin and Kagan \cite{Bur+}. Here we 
have seen that that it can arise without assuming interactions at the level of
quantized excitations, because the statistics of local potential energies is
by itself already a largely collective affair.

\section{Dynamics}\label{sec5}

Let us add a few brief remarks concerning dynamics. We have in mind here
the computation of phonon mean free paths, of internal friction and the
thermal conductivity, the latter two computable once the phonon mean free 
paths are known, as well as the computation of dielectric susceptibilities.
All interesting properties can be computed within linear response theory.
We assume a bilinear interaction of the local coordinates $v_i$,
whose dynamics are given by $H_S = H_{\rm eff}$ in (\ref{Heff}), with the
extended phonon degrees of freedom 
\be
H_B = \sum_{ks} \hbar \omega_{ks} b_{ks}^\dagger b_{ks}
\ee
that is mediated by the strain-field, and of the form
\be
H_{SB} = v \sum_s \gamma_s e_s 
\ee
with $s$ labelling the acoustic branches of the phonon spectrum, $e_s$
denoting the contribution of branch $s$ to the strain field at site $i$
and appropriate coupling constants $\gamma_s$. A Debeye model is assumed
for the phonon bath.

The dynamic properties of interest can be computed from the imaginary part
of the dynamic susceptibility $\chi''_{vv}(\omega)$, which in turn, via 
fluctuation dissipation theorems, is related to the (symmetrized) centered
correlation function
\be
\hat C_{vv}(t) = (\Delta v(t)|\Delta v) = \frac{1}{2}\langle \Delta v(t) 
\Delta v + \Delta v \Delta v(t)\rangle\ ,
\ee
specifically to the spectral function associated with it. Here $\Delta v
= v - \langle v \rangle$.
The Mori-Zwanzig Projection technique and (at this time) a weak coupling
assumption are used to carry through in practice. 

As of now, we have not yet produced quantitative results. So much can,
however, be said at this point. In the limit of very low temperatures,
our model clearly approaches the TLS-physics, and we expect a roughly
quadratic temperature dependence of the thermal conductivity. The increased
density of states in the 1 -- 10\, K range goes along with a new, and
faster scattering mechanism: both are due to the existence of intra--well 
(as opposed to the slower inter--well) transitions in this energy range.
As discussed by Yu and Leggett \cite{YuLe}, the combination of these features
can be expected to produce the plateu in the thermal conductivity.
We shall report on these matters in greater detail in the near future
\cite{HoKu}.

\section{Conclusions and Outlook}\label{sec6}

In summary, we have proposed to take a new look at glassy low--temperature
anomalies from a model--based rather than phenomenological approach. The
models so far considered  do certainly not (yet) attempt to describe the details 
of any specific substance. Our aim has been to formulate something very
schematic that should capture the essentials of amorphous low temperature
phases --- a caricature in the same sense as the Ising model is a caricature
description of uniaxial ferromagnets or, perhaps less boldly, the SK model
a prototypic description of a spin--glass. Chances are of course, that we
still haven't got the bare essentials right. But we feel that our first
attempts do point in the right direction. In particular the clue we have
obtained concerning the understanding as to which phenomena might be 
expected to be univeral and which not, does look like an encouraging 
general qualitative result.

Our approach offers the unique possibility to study the whole range of glassy 
physics, from the regime of the glass transition temperature down to the low 
temperatures where quantum effects play a dominant role, all within a single 
set of model assumptions. In particular, it should be interesting to study 
the glassy dynamics within the type of models we have proposed at or near 
their respective glass transition temperature.

Another issue concerns the classification of spin--glasses into roughly
two families, one exhibiting a continuous transition and infinitely many
levels of RSB, the other a discontinuous transition, and only one level
of RSB, the model class we have presented here belonging to the first.
There are indications that at least from the point of view of dynamics
near the glass transition, the second class should be more appropriate, since 
it exhibits dynamic freezing transitions at temperatures higher than the 
thermodynamic transition temperature seen in equilibrium treatments 
\cite{KiThi+}. We are currently studying a modified version of the Bernasconi 
model \cite{Be87} in this context \cite{DiKu98}. Finally, we have started 
to apply our approach to the case of defect crystals like KCl:Li to study
the effects of the dipolar interaction among the Li defects at higher 
concentration \cite{KuWu98}, a system that has the distinct advantage that
the on-site potential $G(v)$ is well known.

In the present paper we have quantised our system only {\em after\/}
a mean--field decoupling. In principle, the analysis should start out
from a full-fledged quantum-statistical formulation, using imaginary time 
path integrals in conjunction with the replica method. Such a formulation
has been worked out \cite{Ku97} and will be analysed in detail in the near 
future. So much can be said already at this point. Our present approch amounts 
to investigating the quantum--statistical physics in the so--called static
approximation \cite{BrMo80}. In the present context, this approximation 
amounts to ignoring the feedback of quantum fluctuations on the effective 
single site potentials.

{\bf Acknowledgements} This work has been supported by the Deutsche For\-schungsgemeinschaft through the Graduiertenkolleg ``Physikalische Systeme
mit vielen Freiheitsgraden'' (U.H.), and a Heisenberg Fellowship (R.K.).
It is a pleasure to thank C. Enss, H. Horner, 
S. Hunklinger, M. M\'ezard, P. Nalbach,  O. Terzidis,  and A. W\"urger for  
helpful discussions and encouragement. R.K. would like to thank the Physics 
Department of the University of Illinois at Urbana-Champaign for its hospitality
while parts of this paper were being written.

\end{document}